\pdfoutput=1

\documentclass[journal]{IEEEtran}
\usepackage{hyperref}
\usepackage {graphicx,alltt,epsf}
\usepackage{times, amsmath, amssymb, epsfig}
\usepackage{bm}
\usepackage[T1]{fontenc}
\usepackage{multirow}
\usepackage{subfigure}

\begin{document}
%
\title{Improved Design Method for Nearly Linear-Phase IIR Filters Using Constrained Optimization}
\author{R.~C.~Nongpiur,~\IEEEmembership{Member,~IEEE,}
        D.~J.~Shpak,~\IEEEmembership{Senior Member,~IEEE,}
        and~A.~Antoniou,~\IEEEmembership{Life Fellow,~IEEE}

\thanks{Copyright (c) 2012 IEEE. Personal use of this material is permitted. However, permission to use this material for any other purposes must be obtained from the IEEE by sending an email to pubs-permissions@ieee.org.}

\thanks{R.~C.~Nongpiur, D.~J.~Shpak, and A.~Antoniou are with the Department
of Electrical and Computer Engineering, University of Victoria, Victoria,
BC, Canada V8W 3P6 e-mail: rnongpiu@ece.uvic.ca; dshpak@ece.uvic.ca; aantoniou@ieee.org.}

\thanks{Manuscript submitted April 2012.}}

\maketitle

\begin{abstract}
A new optimization method for the design of nearly linear-phase IIR digital filters that satisfy prescribed specifications is proposed. The group-delay deviation is minimized under the constraint that the passband ripple and stopband attenuation are within the prescribed specifications and either a prescribed or an optimized group delay can be achieved. By representing the filter in terms of a cascade of second-order sections, a non-restrictive stability constraint characterized by a set of linear inequality constraints can be incorporated in the optimization algorithm. An additional feature of the method, which is very useful in certain applications, is that it provides the capability of constraining the maximum gain in transition bands to be below a prescribed level. Experimental results show that filters designed using the proposed method have much lower group-delay deviation for the same passband ripple and stopband attenuation when compared with corresponding filters designed with several state-of-the-art competing methods.
\end{abstract}


\section{Introduction}
Linear-phase filters are important in applications where a flat delay characteristic is necessary. Applications of such filters can be found in the field of audio signal processing where it is important to minimize the relative delay between the frequency components to prevent perceptible audio distortion and in digital communications where nonlinear phase can result in signal spreading thereby causing inter-symbol interference between time-concentrated information symbols. Perfectly linear-phase filters can be easily achieved using FIR filters. However, in most applications a perfectly linear phase response is not required and filters that have approximately linear phase response are quite acceptable. In such cases, IIR filters are more attractive than FIR filters for two main reasons. Firstly, they can satisfy the given filter specifications with a much lower filter order thereby reducing the computational requirement and/or the complexity of hardware and, secondly, they have a much smaller group delay.

The presence of the denominator polynomial in IIR filters makes their design more challenging than that of FIR filters because it results in a highly nonlinear objective function that requires highly sophisticated optimization methods. As IIR filters lack the inherent stability of FIR filters, stability constraints must be incorporated in the design process to ensure that the filter is stable, which means constraining the poles to lie within the unit circle of the $z$ plane.

In applications where the phase response is not important, a fairly large choice of methods is available to the filter designer ranging from closed-form methods based on classical analog filter approximations to numerous optimization methods. In~\cite{wsluBook}, unconstrained algorithms of the quasi-Newton family are used in a least-$p$th formulation. Filter stability is achieved by means of a well-known stabilization technique whereby poles outside the unit circle are replaced by their reciprocals. MATLAB\texttrademark function $iirlpnorm$ in~\cite{daleMatlab} implements an unconstrained least-$p$th quasi-Newton algorithm of the type found in~\cite{andreas}. On the other hand, MATLAB\texttrademark function $iirlpnormc$ in~\cite{daleMatlab} implements a least-$p$th Newton method that uses barrier constraints to assure the stability of the resulting filter and also provides for a specified stability margin.

Nearly linear-phase IIR filters can be designed by using the classical $equalizer$ $approach$ whereby an IIR filter satisfying prescribed amplitude-response specifications is designed and is then cascaded with an IIR delay equalizer to approximately linearize the phase response~\cite{andreasPhaseEq}. More recent methods typically involve designing IIR filters that simultaneously satisfy both the amplitude- and phase-response constraints, as it results in filters of lower order~\cite{guindon}-\cite{lai1}. In~\cite{sullivan}, the frequency-response error is minimized under the constraint that the group-delay deviation is within a prescribed level, while in~\cite{guindon} the design problem is formulated as a cascade of filter sections where each section is represented by a biquadratic transfer function either in the conventional polynomial form or in the polar form and the design problem is solved using a constrained Newton method. An important advantage of this method is the nonrestrictive stability constraint that it uses coupled with the capability of controlling the maximum pole radius. Since that objective function is nonconvex, the performance of the algorithm is critically dependent on the initialization point. To overcome difficulties with the nonlinear objective function, iterative methods based on the Steiglitz-McBride (SM) and the Gauss-Newton (GN) algorithms have been employed~\cite{wsluStateOfArt}-\cite{lang}. Although convergence cannot be guaranteed in these methods, solutions that are satisfactory for the intended applications are usually obtained. A drawback of the methods in \cite{wsluStateOfArt}-\cite{lang} is the use of restrictive stability constraints that could prevent better designs.

In~\cite{hinamoto}, a minimax design of IIR filters is formulated as a conic quadratic programming problem by approximating each update as a linear approximation step. An advantage of the method is that it incorporates a nonrestrictive stability constraint using a set of linear inequality constraints; another important advantage is that the final solution is not critically dependent on the initialization point. However, the method has certain drawbacks: it requires the group delay to be specified and, furthermore, it does not provide the option of controlling the relative degree of minimization between the group-delay deviation and the maximum passband ripple.

In this paper, we propose a design method where the group-delay deviation is minimized under the constraints that the passband ripple and stopband attenuation are maintained within the prescribed specifications. The method also provides the option of keeping the group delay fixed at a prescribed value or allowing it to be a free parameter that can be optimized in order to achieve improved filter performance. By designing the filter as a cascade of second-order sections, a necessary and sufficient nonrestrictive stability constraint can be incorporated in the optimization problem as a set of linear inequality constraints.
The method eliminates the drawbacks of the method in~\cite{hinamoto} as it does not require the group delay to be specified, while at the same time it provides the capability of controling the relative quality between the group-delay deviation and the maximum passband ripple. Although the transition region is usually treated as a ``don't care" region, in many practical applications excessive gain in this region could be undesirable. Our method provides the capability of controlling the maximum gain in transition bands. Experimental results show that numerous filters designed with the proposed method have much lower group-delay deviation for the same passband ripple and stopband attenuation than corresponding filters designed with several known competing state-of-the-art methods.

The paper is organized as follows. In Section II, we frame the problem as an iterative constrained optimization problem. Then, in Section III we describe a procedure for designing nearly linear-phase IIR filters. In Section IV, performance comparisons between the filters designed using the proposed method and the existing methods are carried out. Conclusions are drawn in Section V.

\section{The optimization problem}
In this section, we frame the design problem at hand as a constrained optimization problem. To this end, we derive formulations for the stability constraints, group-delay deviation, passband ripple, stopband attenuation, and transition-band gain constraints. Then, we incorporate the formulations within the framework of a constrained optimization problem.

We assume that the filter comprises a cascade of second-order sections (SOSs), which can be represented by a product of biquadratic transfer functions of the form
\begin{equation}
H(\mathbf{c}, z) = H_0 \prod_{m=1}^J \frac{a_{0m}+a_{1m}z+z^2}{b_{0m}+b_{1m}z+z^2}=H_0 \prod_{m=1}^J  \frac{N_m(z)}{D_m(z)}
\label{cascadeSos}
\end{equation}
where
\begin{equation}
\mathbf{c} = [a_{01}~a_{11}~b_{01}~b_{11} \cdots b_{0J}~b_{1J}~H_0]^T
\label{cVec}
\end{equation}
$J$ is the number of filter sections, $N = 2J$ is the filter order, and $H_0$ is a positive multiplier constant. An odd-order transfer function can be readily obtained by setting coefficients $a_{0m}$ and $b_{0m}$ to zero in one SOS.

\subsection{Group-delay deviation}
The group delay corresponding to transfer function $H(\mathbf{c}, z)$ in (\ref{cascadeSos}) is given by~\cite{andreasPhaseEq}
\begin{equation}
\tau_h(\mathbf{c}, z) = -\sum_{i=1}^J \frac{\alpha_n(\mathbf{c}, z, i)}{\beta_n(\mathbf{c}, z, i)} + \sum_{i=1}^J \frac{\alpha_d(\mathbf{c}, z, i)}{\beta_d(\mathbf{c}, z, i)}
\end{equation}
where
\begin{eqnarray}
\alpha_n(\mathbf{c}, z, i) & = & 1 - a_{0i}^2 + a_{1i}(1-a_{0i})\cos\omega \\
\beta_n(\mathbf{c}, z, i) & = & a_{0i}^2+a_{1i}^2+1+ 2a_{0i}(2\cos^2\omega-1) \\
& & + 2a_{1i}(a_{0i}+1)\cos\omega \notag \\
\alpha_d(\mathbf{c}, z, i) & = & 1- b_{0i}^2+b_{1i}(1-b_{0i}) \cos\omega \\
\beta_d(\mathbf{c}, z, i) & = & b_{0i}^2+b_{1i}^2+1+2b_{0i}(2\cos^2\omega-1) \\
& & +2b_{1i}(b_{0i}+1)\cos\omega \notag
\end{eqnarray}
The group-delay deviation at frequency $\omega$ is given by
\begin{equation}
e_g(\mathbf{x}, e^{j\omega}) = \tau_h(\mathbf{c}, e^{j\omega}) - \tau
\end{equation}
where
\begin{equation}
\mathbf{x} = [\mathbf{c}^T \tau]^T
\end{equation}
and $\tau$ is the desired group delay, which may be an optimization variable. If $\mathbf{x}_k$ is the value of $\mathbf{x}$ at the start of the $k$th iteration and ${\bm \delta}$ is the update to $\mathbf{x}_k$, the updated value of the group-delay deviation can be estimated by a linear approximation
\begin{equation}
e_g(\mathbf{x}_k+{\bm \delta}, e^{j\omega}) \approx e_g(\mathbf{x}_k, e^{j\omega}) + \nabla e_g(\mathbf{x}_k, e^{j\omega})^T {\bm \delta}
\label{gdError}
\end{equation}
which becomes more accurate as $\|{\bm \delta} \|_2$ gets smaller.

If $\omega_{pl}$ and $\omega_{ph}$ are the lower and upper edges of the passband, the $L_p$-norm of the passband group-delay deviation for the $k$th iteration is given by
\begin{eqnarray}
\mathbf{E}_p^{(gd)}(k) & = &  \left[ \int_{\omega_{pl}}^{\omega_{ph}} | e_g(\mathbf{x}_{k+1}, e^{j\omega})|^p d\omega \right]^{1/p} \notag \\
& \approx & \kappa_{g} \left[ \sum_{i=1}^N  |e_g(\mathbf{x}_{k+1}, e^{j\omega_i})|^p \right]^{1/p}, ~\omega_i \in \Psi_p \notag \\
& \approx & \left[ \sum_{i=1}^N |\kappa_{g} e_g(\mathbf{x}_k, e^{j\omega_i}) + \right. \notag \\
& &  \ \ \ \ \kappa_{g} \nabla e_g(\mathbf{x}_k, e^{j\omega_i})^T {\bm \delta}|^p \Big]^{1/p} \notag \\
\label{gdDev}
\end{eqnarray}
where $\Psi_p \in [\omega_{pl}, \omega_{ph}]$ is the set of passband frequency sample points and $\kappa_{g}$ is a constant. Expressing (\ref{gdDev}) in matrix form, we get
\begin{eqnarray}
\mathbf{E}_p^{(gd)}(k) & \approx & \|\mathbf{C}_k {\bm \delta} + \mathbf{d}_k \|_p  \label{gdDevMat}
\end{eqnarray}
where
\begin{eqnarray}
\mathbf{C}_k &=& \left[ \begin{matrix}
\kappa_{g} \nabla e_g(\mathbf{x}_k, e^{j\omega_1})^T  \cr
\vdots   \cr
\kappa_{g} \nabla e_g(\mathbf{x}_k, e^{j\omega_{N_p}})^T  \cr
\end{matrix}
\right ]  \\
\mathbf{d}_k &=& [d_1~d_2~\cdots~d_{N_p}]^T, \\
d_i & = & \kappa_{g} e_g(\mathbf{x}_k, e^{j\omega_i}), \ \  \omega_i \in \Psi_p
\label{C_gd}
\end{eqnarray}
The right-hand side of (\ref{gdDevMat}) is the $L_p$-norm of an affine function of ${\bm \delta}$ and, therefore, it is convex with respect to ${\bm \delta}$~\cite{wsluBook}.

The quality of the group-delay characteristic of the filter can be measured by using the normalized maximum variation of the filter group delay, $\tau_h$, over the passband as a percentage, i.e.,~\cite{andreas}
\begin{equation}
Q_{\tau} = \frac{100(\tau_{max} -\tau_{min})}{2\tau_{avg}}
\end{equation}
where
\begin{eqnarray}
\tau_{avg} & = & \frac{\tau_{max}+\tau_{min}}{2} \\
\tau_{max} & = & \max_{\omega \in \Psi_p} \tau_h(e^{j\omega}) \\
\tau_{min} & = & \min_{\omega \in \Psi_p}  \tau_h(e^{j\omega})
\end{eqnarray}
Hence,
\begin{equation}
Q_{\tau} = \frac{100(\tau_{max} -\tau_{min})}{(\tau_{max}+\tau_{min})}
\label{Qval}
\end{equation}
$Q_{\tau}$ will be referred to as the {\it maximum group-delay deviation} hereafter.

\subsection{Passband error}
If $H_d(\omega)$ is the desired frequency response of the filter and $\mathbf{c}_k$ is the value of vector $\mathbf{c}$ at the start of the $k$th iteration, a passband error function at frequency $\omega$ can be defined as
\begin{equation}
e_h(\mathbf{c}_k, e^{j\omega}) = |H(\mathbf{c}_k, e^{j\omega})|^2 - |H_d(\omega)|^2
\end{equation}
Without loss of generality, we can assume that the desired amplitude response is unity in the passband. Therefore, the passband error function becomes
\begin{equation}
e_h^{(pb)}(\mathbf{c}_k, e^{j\omega}) = |H(\mathbf{c}_k, e^{j\omega})|^2 - 1, \ \omega \in \Psi_p
\end{equation}
Using the same approach as in Section II-A, the $L_p$-norm of the passband error function, $\mathbf{E}_p^{(pb)}(k)$, in matrix form can be approximated as
\begin{equation}
\mathbf{E}_p^{(pb)}(k) \approx  \|\mathbf{D}_k^{(pb)} {\bm \delta} + \mathbf{f}_k^{(pb)} \|_p
\label{pbConstr}
\end{equation}
where
\begin{eqnarray}
\mathbf{D}_k^{(pb)} &=& \left[ \begin{matrix}
\kappa_{pb} \nabla e_h^{(pb)}(\mathbf{c}_k, e^{j\omega_1})^T & 0 \cr
\vdots  & \vdots \cr
\kappa_{pb} \nabla e_h^{(pb)}(\mathbf{c}_k, e^{j\omega_{N_p}})^T & 0\cr
\end{matrix}
\right ], \ \  \omega_i \in \Psi_p \label{Dmat_pb} \nonumber\\
\\
\mathbf{f}_k^{(pb)} &=& [f_1^{(pb)}~f_2^{(pb)}~\cdots ~f_{N_p}^{(pb)}]^T, \\
f_i^{(pb)} & = & \kappa_{pb} e_h^{(pb)}(\mathbf{c}_k, e^{j\omega_i}), \\
{\bm \delta} & = & [{\bm \delta}_c^T \delta_{\tau}]^T
\label{vectorUpdate}
\end{eqnarray}
In the above equations, ${\bm \delta}_c$ is the vector update for $\mathbf{c}_k$, $\delta_{\tau}$ is the scalar update for $\tau$, and $\kappa_{pb}$ is a constant. The elements of the last column of $\mathbf{D}_k^{(pb)}$ in (\ref{Dmat_pb}) are all zeros since (\ref{pbConstr}) is independent of $\tau$.

\subsection{Amplitude response in the stopband and transition band}
The frequency response update for the filter at the $k$th iteration is given by
\begin{equation}
H(\mathbf{c}_k+{\bm \delta}_c, e^{j\omega}) \approx  H(\mathbf{c}_k, e^{j\omega}) + \nabla H(\mathbf{c}_k, e^{j\omega})^T {\bm \delta}_c
\end{equation}
By using the same approach as in Section II-B, the $L_p$-norm of the frequency response in the stopband and transition band can be approximated as
\begin{eqnarray}
\mathbf{E}_p^{(sb)}(k) & \approx &  \|\mathbf{D}_k^{(sb)} {\bm \delta} + \mathbf{f}_k^{(sb)} \|_p  \label{sbConstr} \\
\mathbf{E}_p^{(tb)}(k) & \approx & \|\mathbf{D}_k^{(tb)} {\bm \delta} + \mathbf{f}_k^{(tb)} \|_p  \label{tbConstr}
\end{eqnarray}
where
\begin{eqnarray}
\mathbf{D}_k^{(sb)} &=& \left[ \begin{matrix}
\kappa_{sb} \nabla H(\mathbf{c}_k, e^{j\omega_1})^T & 0 \cr
\vdots  & \vdots \cr
\kappa_{sb} \nabla H(\mathbf{c}_k, e^{j\omega_{N_s}})^T & 0\cr
\end{matrix}
\right ], \ \  \omega_i \in \Psi_s  \\
\mathbf{D}_k^{(tb)} &=& \left[ \begin{matrix}
\kappa_{tb} \nabla H(\mathbf{c}_k, e^{j\omega_1})^T & 0 \cr
\vdots  & \vdots \cr
\kappa_{tb} \nabla H(\mathbf{c}_k, e^{j\omega_{N_t}})^T & 0\cr
\end{matrix}
\right ], \ \  \omega_i \in \Psi_t  \\
\mathbf{f}_k^{(sb)} &=& [f_1^{(sb)}~f_2^{(sb)}~\cdots~f_{N_s}^{(sb)}]^T, \ \  \omega_i \in \Psi_s \\
\mathbf{f}_k^{(tb)} &=& [f_1^{(tb)}~f_2^{(tb)}~\cdots~f_{N_t}^{(tb)}]^T, \ \  \omega_i \in \Psi_t \\
f_i^{(sb)} & = & \kappa_{sb} H(\mathbf{c}_k, e^{j\omega_i}) \\
f_i^{(tb)} & = & \kappa_{tb} H(\mathbf{c}_k, e^{j\omega_i})
\label{D_ampResp}
\end{eqnarray}
In the above equations, $\Psi_s$ and $\Psi_t$ are the sets of frequency sample points in the stopband and transition band, respectively, and $\kappa_{sb}$ and $\kappa_{tb}$ are constants.  Note that $\Psi_p$, $\Psi_s$, and $\Psi_t$ may extend over two or more passbands, stopbands, or transition bands, respectively, in the case of multiband filters.

\subsection{Filter stability}
To ensure that the filter is stable, the poles of the transfer function must lie within the unit circle~\cite{andreas}. If $\epsilon_s > 0$ is a stability margin defined as $\epsilon_s =1-r_{max}^{(p)}$ where $r_{max}^{(p)}$ is the maximum pole radius allowed, the coefficients of the denominator should satisfy the stability conditions given by~\cite{guindon}
\begin{eqnarray}
b_{0m} & \leq & 1-\gamma \\
b_{1m} - b_{0m} & \leq & 1-\gamma \\
-b_{1m} - b_{0m} & \leq & 1-\gamma
\end{eqnarray}
where~\cite{hinamoto}
\begin{equation}
\gamma = 1-(1-\epsilon_s )^2
\label{stabMargin}
\end{equation}
Incorporating stability margin $\epsilon_s$ would ensure that roundoff errors would not cause filter instability particularly in fixed-point implementations.

The stability conditions in (37)-(39) for the $k$th iteration can be expressed in matrix form as
\begin{equation}
\mathbf{B}{\bm \delta} < \mathbf{b}^{(k)}
\label{stabConstr}
\end{equation}
where
\begin{eqnarray}
\mathbf{B} &=& \left[ \begin{matrix}
\mathbf{\mathcal{B}} & \mathbf{0} & \cdots & \mathbf{0} & 0\ 0 \cr
\vdots  & \vdots & \ddots & \vdots & \vdots \cr
\mathbf{0} & \mathbf{0} & \cdots & \mathbf{\mathcal{B}} & 0\ 0 \cr
\end{matrix} \right] \\
\mathbf{\mathcal{B}} &=& \left[ \begin{matrix}
1 & \ \ 0 \cr
-1 & \ \ 1 \cr
-1 & -1 \cr
\end{matrix} \right] \\
\mathbf{b}^{(k)} &=& \left[ \begin{matrix}
\mathbf{b}_1^{(k)} \cr
\vdots \cr
\mathbf{b}_J^{(k)} \cr
\end{matrix} \right] \\
\mathbf{b}_m^{(k)} &=& \left[ \begin{matrix}
1-\epsilon_s - b_{0m}^{(k)}  \cr
1 - \epsilon_s - b_{1m}^{(k)} + b_{0m}^{(k)} \cr
1 - \epsilon_s + b_{1m}^{(k)} + b_{0m}^{(k)} \cr
\end{matrix} \right]
\end{eqnarray}

\subsection{Optimization problem}
The optimization is carried out by minimizing the group-delay deviation under the constraints that the passband error, stopband attenuation, and transition band attenuation are within prescribed levels. This can be done by solving the optimization problem
\begin{eqnarray}
\mbox{minimize } & &  \|e_g(\mathbf{x}, e^{j\omega})\|_p \label{optProblem}   \\
\mbox{subject to: } & & \mbox{passband error function} < \Gamma_{pb} \notag \\
& & \mbox{stopband gain } < \Gamma_{sb} \notag  \\
& & \mbox{transition band gain } < \Gamma_{tb} \notag \\
& & \mbox{stability margin$~ = ~\epsilon_s$} \notag
\end{eqnarray}
where $\Gamma_{pb}$, $\Gamma_{sb}$, and $\Gamma_{tb}$ are the maximum prescribed levels for the
passband error, stopband gain, and transition band gain, respectively.

Using (\ref{gdDevMat}), (\ref{pbConstr}), (\ref{sbConstr}), (\ref{tbConstr}), and (\ref{stabConstr}) the problem to be solved in the $k$th iteration can be expressed as
\begin{eqnarray}
\mbox{minimize } & &  \|\mathbf{C}_k {\bm \delta} + \mathbf{d}_k \|_p  \label{optProbConvex} \\
\mbox{subject to: } & &
\|\mathbf{D}_k^{(pb)} {\bm \delta} + \mathbf{f}_k^{(pb)} \|_p < \Gamma_{pb} \notag  \\
& & \|\mathbf{D}_k^{(sb)} {\bm \delta} + \mathbf{f}_k^{(sb)} \|_p < \Gamma_{sb} \notag  \\
& & \|\mathbf{D}_k^{(tb)} {\bm \delta} + \mathbf{f}_k^{(tb)} \|_p < \Gamma_{tb} \notag \\
& & \|{\bm \delta}\|_2 < \Gamma_{small} \notag \\
& & \mathbf{B}{\bm \delta} < \mathbf{b}^{(k)} \notag
\end{eqnarray}
where ${\bm \delta}$ is the optimization variable. The optimum value of ${\bm \delta}$ is then used to update the optimizing parameters for the next iteration. The bound on the norm of the update vector assures the validity of the linear approximation and at the same time eliminates the need for a line search step that is required in most optimization algorithms~\cite{hinamoto}.

The value of $p$ in the above optimization problem can be any positive integer.  The most significant values for $p$ are 2 and $\infty$.  In the first case, the $L_2$-norm would be minimized, which would result in a least-squares solution, whereas, in the second case, the $L_\infty$-norm would be minimized, which would result in a minimax solution.  In a least-squares solution, the squares of the passband and stoband errors would be minimized whereas in a minimax solution the maxima of the absolute values of the passband and stopband errors would be
minimized. In this paper, we explore the use of the $L_\infty$-norm.

The optimization problem in (\ref{optProbConvex}) can be solved by starting with an initialization filter that satisfies at least some of the specifications, if at all possible, in order to reduce the amount of computation required.  The initialization filter may satisfy only the stopband and transition band constraints but not the passband constraint.
In order to handle such scenarios, we relax the constraints for the passband and the norm of the
parameter update by adding a variable $\delta_{rlx}$ which is also minimized while its value is constrained to be
positive; when $\delta_{rlx} = 0$, the original constraints are restored. The relaxation of the norm of the parameter update is to facilitate a greater change between iterations so that all constraints are quickly satisfied and $\delta_{rlx}$ is reduced to a small value. With this modification, the problem in (\ref{optProbConvex}) for the case $p=\infty$ becomes
 \begin{eqnarray}
\mbox{minimize } & &  \|\mathbf{C}_k {\bm \delta} + \mathbf{d}_k \|_\infty + W\delta_{rlx} \label{optProbConvexMod} \\
\mbox{subject to: } & &
\|\mathbf{D}_k^{(pb)} {\bm \delta} + \mathbf{f}_k^{(pb)} \|_\infty < \Gamma_{pb} + \delta_{rlx} \notag  \\
& & \|\mathbf{D}_k^{(sb)} {\bm \delta} + \mathbf{f}_k^{(sb)} \|_\infty < \Gamma_{sb} \notag  \\
& & \|\mathbf{D}_k^{(tb)} {\bm \delta} + \mathbf{f}_k^{(tb)} \|_\infty < \Gamma_{tb} \notag \\
& & \|{\bm \delta}\|_2 < \Gamma_{small}+\delta_{rlx} \notag \\
& & \delta_{rlx} > 0 \notag \\
& & \mathbf{B}{\bm \delta} < \mathbf{b}^{(k)}  \notag
\end{eqnarray}
where ${\bm \delta}$ and $\delta_{rlx}$ are optimization variables, and $W>0$ is a weighing factor for the relaxation variable. It should be noted that if the initialization filter is feasible, the use of the relaxation parameter $\delta_{rlx}$ is not necessary although in some cases it has been found to speed up the convergence. For the situation where the numerator and denominator orders are required to be different, the appropriate numerator or denominator coefficients in some of the SOSs can be set to zero by initializing them with zero and correspondingly constraining their updated values to zero. That is, if $x_k(i)$ corresponds to a coefficient in an SOS that is required to be zero in every iteration $k$, we intialize $x_0(i)=0$ and correspondingly include the additional constraint on the corresponding updated value, i.e., $\delta(i) = 0$.

The optimization problem in (\ref{optProbConvexMod}) can be easily expressed as a {\it second-order cone programmming} (SOCP) problem as in~\cite{hinamoto} and solved using efficient SOCP solvers such as the one available in the SeDuMi optimization toolbox~\cite{sedumi} for MATLAB\texttrademark. The various $L_\infty$-norms in (\ref{optProbConvexMod}) are efficiently evaluated by the SOCP solver.

\section{Design procedure}
Two general strategies for the design of digital filters have been developed to deal with design problems where the group delay is not specified or with problems where a prescribed group delay is required.  In the former case, the group delay can be used as an additional independent variable that can be optimized in order to bring about additional improvements to the filter being designed.

\subsection{Optimized group delay}
When the group delay is assumed to be an independent variable, it is important that the initialization filter be chosen to be close to the desired optimal filter in order to assure fast convergence. To this end, a good first step would be to design the lowest-order IIR filter that satisfy only the amplitude-response specifications. An {\it elliptic} filter would be the most suitable choice since it gives the lowest-order IIR filter for any given amplitude-response specifications because of the optimality of the elliptic approximation. Such a filter can be obtained by using the design method described in Chap.~12 of~\cite{andreas}.

To reduce the group-delay deviation of the filter in the passband, a number of additional general biquadratic SOSs are included depending on the degree of linearity required in the phase response.  To achieve fast convergence, the additional SOSs are initialized as allpass sections. The poles and zeros of the additional SOSs are initially distributed in the passband sector of the $z$ plane, namely, the sector bounded by the passband edge frequencies. Under these circumstances, the transfer function assumes the form
\begin{equation}
\begin{split}
H_{init}(z) & = \ H_{ellip}(z)  \\
& \cdot G_0 \prod_{k=1}^{M} \frac{(z-r_k^{-1}e^{j\omega_k})(z-r_k^{-1}e^{-j\omega_k})}{(z-r_k e^{j\omega_k})(z-r_k e^{-j\omega_k})}, \ \ \ \omega_k \in \Psi_p
\end{split}
\label{initFilter}
\end{equation}
where $H_{ellip}(z)$ is the transfer function of the elliptic filter, $G_0$ is a normalizing gain factor, $M$ is the number of additional allpass SOSs, and $0 < r_k < 1$. An initialization group delay that was found to work well is the average of the passband filter-equalizer combination, which can be estimated as
\begin{equation}
\tau_{init} = \frac{-1}{\omega_{ph}-\omega_{pl}} \int_{\omega_{pl}}^{\omega_{ph}} \frac{d}{d\omega}\left\{ \mbox{arg} \left[ H_{init}(e^{j\omega})\right]\right\} d\omega
\label{initTau}
\end{equation}
Although using the initializaton group delay in (\ref{initTau}) always results in solutions that are comparable or better than those achieved with existing methods, we have found some instances where starting with the maximum or minimum passband group delay can yield a better solution. Therefore, to maximize the possibility for obtaining the best solution, the optimization can also be initialized with the minimum and maximum passband group-delays given by
\begin{equation}
\tau_{init}^{(max)} =  \max_\omega \left(- \frac{d}{d\omega}\left\{ \mbox{arg} \left[ H_{init}(e^{j\omega})\right] \right\} \right), \ \omega \in \Psi_p \\
\end{equation}
\begin{equation}
\tau_{init}^{(min)}  =  \min_\omega \left(- \frac{d}{d\omega}\left\{ \mbox{arg} \left[ H_{init}(e^{j\omega})\right] \right\} \right),\ \omega \in \Psi_p
\end{equation}
and then selecting the best solution from among the three results. However, if there is a need to reduce the amount of computation required, the filter can be designed by using only the initialization group delay in (\ref{initTau}).

The required filter can be designed by using the following algorithm:
\subsubsection*{Step 1}
Obtain the transfer function of the required elliptic filter, $H_{ellip}(z)$, that satisfies the required amplitude response specifications; e.g., by using the D-Filter software package~\cite{dfilter}.
\subsubsection*{Step 2}
Set the number of additional general biquadratic SOSs to $M$ and select $r_k$ and $\omega_k$ to construct the transfer function $H_{init}(z)$ as in (\ref{initFilter}); from $H_{init}(z)$, compute the initialization group delay $\tau_{init}$ using (\ref{initTau}). This can be easily done by using D-Filter~\cite{dfilter}.
\subsubsection*{Step 3a}
Using $H_{init}(z)$ and $\tau_{init}$ for initialization, solve the optimization problem in (\ref{optProbConvexMod}).
\subsubsection*{Step 3b (optional)}
 Solve the optimization problem in (\ref{optProbConvexMod}) using $\{H_{init}(z), \tau_{init}^{(max)}\}$ and $\{H_{init}(z), \tau_{init}^{(min)}\}$ for initialization and  then select the solution that has the smallest value of $Q_\tau$ in {\it Steps 3a} and {\it 3b}.
\subsubsection*{Step 4}
Using (\ref{Qval}), compute the maximum group-delay deviation of the filter, $Q_{\tau}$, obtained in {\it Step 3}. If $Q_{\tau}$ is less than the prescribed value, the filter specifications are satisfied and the algorithm is terminated; otherwise, set $M=M+1$ and go to {\it Step 2}.

The optional step, {\it Step 3b}, can be carried out if the amount of computation required is not a critical factor, in order to increase the possibility for obtaining a better solution.

\subsection{Prescribed group delay}
When a prescribed group delay is required, the initialization procedure described in Sec.~III-A is not appropriate. A more appropriate initialization scheme would be to use the balance model truncation (BMT) method described in~\cite{kaleIeee, pernebo}. The main steps of the BMT involve converting a high-order FIR filter into a state-space balanced model, then reducing the model order, and finally converting the lower-order model to a reduced-order IIR filter.

To ensure that the IIR filter obtained with the BMT method has a group delay that is close to the prescribed value, the initialization linear-phase FIR filter is designed to have a group delay, $\tau_{pr}$, that is close to the prescribed value.  This can be done by selecting the filter length as
\begin{equation}
L_{fir} = 2\lceil \tau_{pr} \rceil + 1
\label{bmtLen}
\end{equation}
where $\lceil \cdot \rceil$ is the ceiling operator. The transfer function of the IIR filter can be expressed as
\begin{equation}
H_{init}(z) = \hat{G}_0 \frac{\displaystyle \prod_i (z - \check{z}_i)}{{D_{bmt}(z)}} =  \frac{\displaystyle \sum_i \check{b}_i z^{m-i}}{D_{bmt}(z)}
\label{numTruncate}
\end{equation}
where the normalizing gain factor $\hat{G}_0$ is chosen to ensure that the average passband gain of the filter is unity.

Sometimes, the filter obtained with the BMT method may have one or more zeros that are located far away from the origin. Such zeros can slow down the optimization algorithm. Experimental results have shown that faster convergence can be achieved by moving any zeros with a radius greater than a prescribed maximum $r_{max}^{(z)}$ to the origin of the $z$ plane by letting
\begin{equation}
\check{z}_i = \begin{cases} 0 & \mbox{ if $|\bar{z}_i| > r_{max}^{(z)}$ } \\
\bar{z}_i & \mbox{ otherwise}
\end{cases}
\label{rmax}
\end{equation}

With the group delay fixed, the problem in (\ref{optProbConvexMod}) can be simplified to
  \begin{eqnarray}
\mbox{minimize } & &  \|\mathbf{C}_k \mathbf{I}_{ce} {\bm \delta}_c + \mathbf{d}_k \|_\infty + W\delta_{rlx} \label{optProbConvexFixedTau} \\
\mbox{subject to: } & &
\|\mathbf{D}_k^{(pb)} \mathbf{I}_{ce} {\bm \delta}_c + \mathbf{f}_k^{(pb)} \|_\infty < \Gamma_{pb} + \delta_{rlx} \notag  \\
& & \|\mathbf{D}_k^{(sb)} \mathbf{I}_{ce} {\bm \delta}_c + \mathbf{f}_k^{(sb)} \|_\infty < \Gamma_{sb} \notag  \\
& & \|\mathbf{D}_k^{(tb)} \mathbf{I}_{ce} {\bm \delta}_c + \mathbf{f}_k^{(tb)} \|_\infty < \Gamma_{tb} \notag \\
& & \|{\bm \delta}_c\|_2 < \Gamma_{small}+\delta_{rlx} \notag \\
& & \delta_{rlx} > 0 \notag \\
& & \mathbf{B} \mathbf{I}_{ce} {\bm \delta}_c < \mathbf{b}^{(k)} \notag
\end{eqnarray}
where ${\bm \delta}_c$ is the optimization variable and $\mathbf{I}_{ce}$ is given by
\begin{equation}
\mathbf{I}_{ce} = \left[ \begin{matrix}
1 & 0 & \cdots  & 0 \cr
0 & 1 & \cdots  & 0 \cr
\vdots & \vdots & \ddots & \vdots \cr
0 & 0 & \cdots & 1 \cr
0 & 0 & \cdots & 0 \cr
\end{matrix} \right] \\
\end{equation}
The required filter can then be designed by using the following procedure:
\subsubsection*{Step 1}
Design a linear-phase FIR filter of length given by (\ref{bmtLen}) with the prescribed passband- and stopband-edge frequencies, by using D-Filter or the MATLAB\texttrademark function {\it fir1} or some other way.
\subsubsection*{Step 2}
If the total number of SOSs is $M_{tot}$, the IIR filter order is $2M_{tot}$. Using the BMT method, transform the FIR filter obtained in {\it Step 1} to an IIR filter of order $2M_{tot}$.
\subsubsection*{Step 3}
Form the transfer function in (\ref{numTruncate}).
\subsubsection*{Step 4}
Using $\check{H}_{init}(z)$ in (\ref{numTruncate}) for initialization, solve the optimization problem in (\ref{optProbConvexFixedTau}) for the prescribed group delay of $\tau_{pr}$.
\subsubsection*{Step 5}
Using (\ref{Qval}) compute the group-delay deviation, $Q_{\tau}$, of the filter obtained from {\it Step 4}. If $Q_{\tau}$ is less than the maximum prescribed value, the filter specifications are satisfied and the algorithm is terminated; otherwise, set $M_{tot}=M_{tot}+1$ and go to {\it Step 2}.

\subsection{Practical considerations}
The computational effort required to complete a design is directly proportional to the number of frequency sample points used in sets $\Psi_p$, $\Psi_s$, and $\Psi_t$.  To reduce the number of sample points and at the same time prevent spikes in the error function, the {\it nonuniform variable sampling} (NVS) technique described in~\cite{andreasNonSampl} (see also Chapter 16 of~\cite{andreas}) was used for the choice of sample points. When compared with the standard uniform sampling method, the NVS technique not only reduces the computational effort by around an order of magnitude but it also reduces the passband ripple and improves the convergence of our algorithms.

The weight $W$ for the relaxation parameter, $\delta_{rlx}$, in (\ref{optProbConvexMod}) should not be too small as this can make the optimization algorithm unstable and prevent it from converging; at the same time, it should also not be too large as this can slow down the convergence process. Typical values of $W$ that have been found to work well range between 500 to 5000.

The computational efficiency of the algorithm is also dependent on two other aspects: the efficiency of the SOCP solver and the number of iterations required to achieve convergence. For the former, most SOCP solvers, including the one in the SeDuMi toolbox for MATLAB\texttrademark, are polynomial-time type algorithms which can be easily run on inexpensive computers such as a laptop. For the latter, the number of iterations to achieve convergence is determined by how close the initialization point is to the final solution. Sometimes, the solution keeps improving with each iteration, but beyond a certain point the degree of improvement is too small to be of practical significance and the optimization can be terminated. Sometimes, the objective function may at some point show very small improvement, or even increase for several iterations before rapidly decreasing again.  To ensure that the optimization is not prematurely terminated, the values of the objective function for the $L_o$ most recent iterations are compared with the minimum value achieved before and if each of these values is larger than the minimum value, the optimization is terminated. A value of 40 for $L_o$ was found to work well.

\section{Experimental Results}
In order to compare our method with other state-of-the-art competing methods, we have designed and tested many nearly linear-phase IIR filters satisfying a diverse range of specifications. Ten examples are included in this paper.

Parameters $\Gamma_{small}$ and $W$ in (\ref{optProbConvexMod}) were set to 0.01 and 1000, respectively, while $r_{max}^{(z)}$ in (\ref{rmax}) was set to 2.5.  The $M$ poles of the allpass part of the transfer function in (\ref{initFilter}) were initialized with $r_k = 0.8$ and $\omega_k$ uniformly distributed in the passband of the required filter. The default maximum pole radius was set to 0.98 except in Example 10 where it was set to a value slightly higher than the maximum pole radius of the elliptic filter, namely, 0.991.

The frequency sets $\Psi_p$, $\Psi_s$, and $\Psi_t$ were determined by using the NVS technique  \cite{andreasNonSampl} with a dense set of 2000 uniformly-spaced virtual frequencies and 68 actual sample frequencies for each passband and stopband, respectively. Six of the actual sample frequencies located near the passband edge in $\Psi_p$ and stopband edge in $\Psi_s$ were fixed with a separation of $7.8\times 10^{-4}$ radians between them.  For each transition band, 500 virtual sample frequencies and 18 actual sample frequencies were used.

In some of the design examples, namely, in Examples 1, 2, and 3, the passband gain of the competing filter used for comparison was constrained to have a maximum value of 1. To achieve consistency between our filters and the competing filters, we normalized the multiplier constant in some of the competing designs to achieve an average passband gain of unity. This was done by adjusting the multiplier constant so that
\begin{equation}
\bar{H}_0 = \frac{2H_0}{g_{max}+g_{min}} = \frac{2H_0}{1+g_{min}}
\end{equation}
where $g_{max}$ and $g_{min}$ are the maximum and minimum passband gains, respectively.

In the experiments, the optimization was first carried out by allowing the group delay to be variable without imposing a constraint on the maximum gain in transition bands. In some designs, the filter obtained had an anomaly in a transition band whereby the gain increased above unity. This may not be a problem in many applications but if the anomaly is undesirable it can be eliminated by imposing a constraint on the maximum gain for the transition band with the anomaly. In such situations, a second design was carried out to ensure that the gain in transition bands is always equal to or less than unity.

Our method can be used to design filters with optimized or fixed group delay. However, some of the competing methods do not provide for an optimized group delay. In such examples, we have designed two filters, one with fixed and the other with optimized group delay for the sake of comparison.

To deal with the above possibilities, some of the following four design variants have been carried out for the examples considered:
\begin{enumerate}
\item{\it Design A-1} The group delay is optimized and no constraint is applied on the transition-band gain.
\item{\it Design A-2} The group delay is optimized and a constraint is applied on the transition-band gain with $\Gamma_{tb} = 1$.
\item{\it Design B-1} The group delay is fixed and no constraint is applied on the transition-band gain.
\item{\it Design B-2} The group delay is fixed and a constraint is applied on the transition-band gain with $\Gamma_{tb} = 1$.
\item{\it Design B-3} The group delay is fixed and $\Gamma_{tb}$ is adjusted so as to ensure that the transition band gain is close to that of the competing filter.
\end{enumerate}

Examples 1 to 3 involve a lowpass, a highpass, and a bandpass filter and the designs obtained are compared with designs based on the classical equalizer approach described in~\cite{andreasPhaseEq}. Examples 4, 5, 6, 8, and 9 concern different lowpass filters and Example 7 concerns a highpass filter. The designs obtained with the proposed method are compared with corresponding designs obtained by using the methods in~\cite{guindon}, \cite{hinamoto}, \cite{lang}, \cite{jiang}, \cite{jiang2}, \cite{lai1}, \cite{sullivan}, and \cite{lai2}, respectively. Example 10 is concerned with a selective IIR bandpass-filter design which is compared with an optimal FIR filter satisfying the same specifications.

It should be noted that the optional design step, {\it Step 3b}, in Section III-A for the design of filters with optimized group delay was included in all of the design examples except Example 10.

\subsection{Example 1, 2, and 3}
\begin{figure}
\begin{center}
\includegraphics[width=0.48\textwidth]{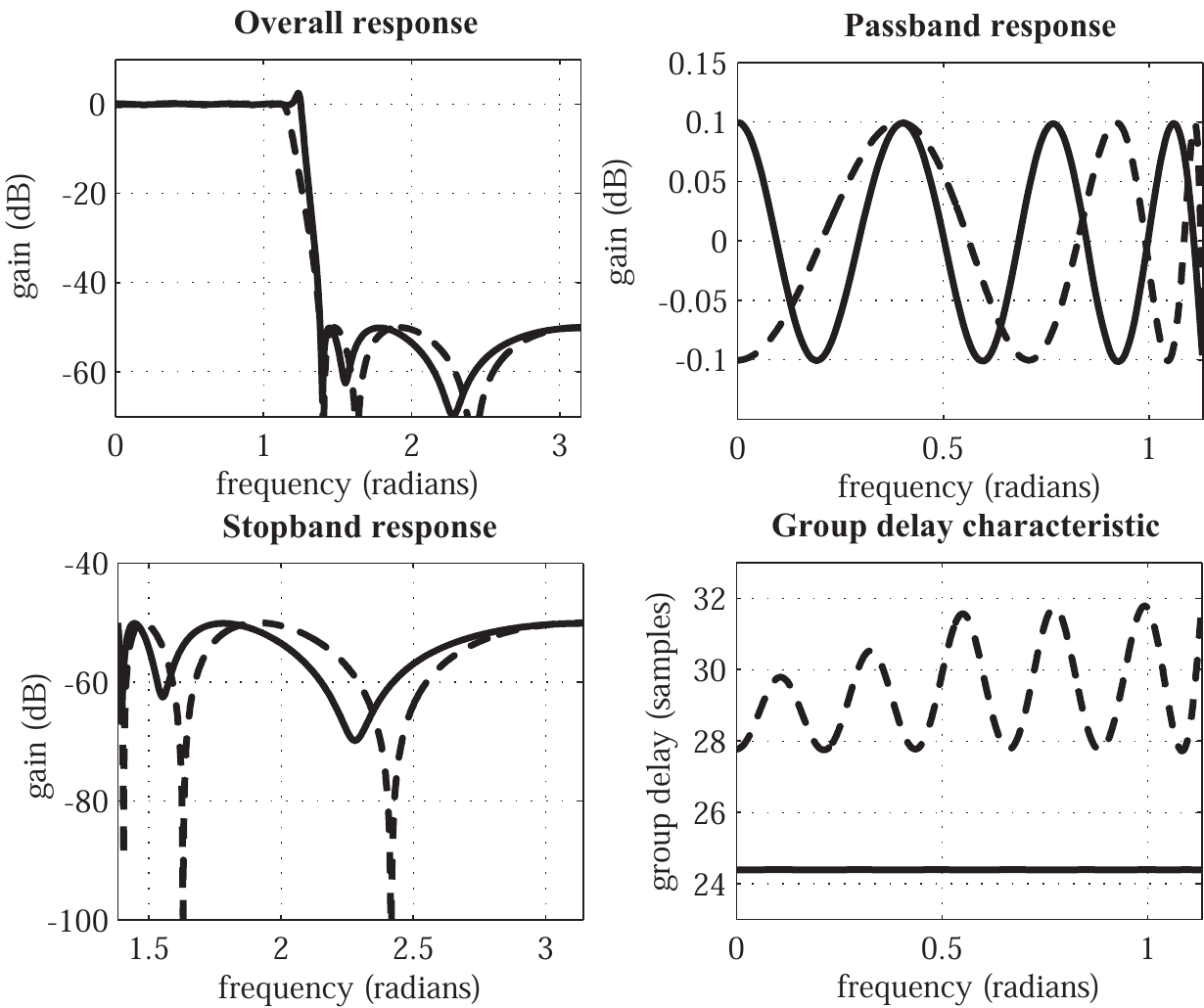}
\caption{Overall, passband, and stopband amplitude responses and group-delay characteristic for Design A-1 of the proposed method (solid curves) and the method in~\cite{andreasPhaseEq} (dashed curves) for Example 1.}
\label{eg1}
\end{center}
\end{figure}

\begin{figure}
\begin{center}
\includegraphics[width=0.48\textwidth]{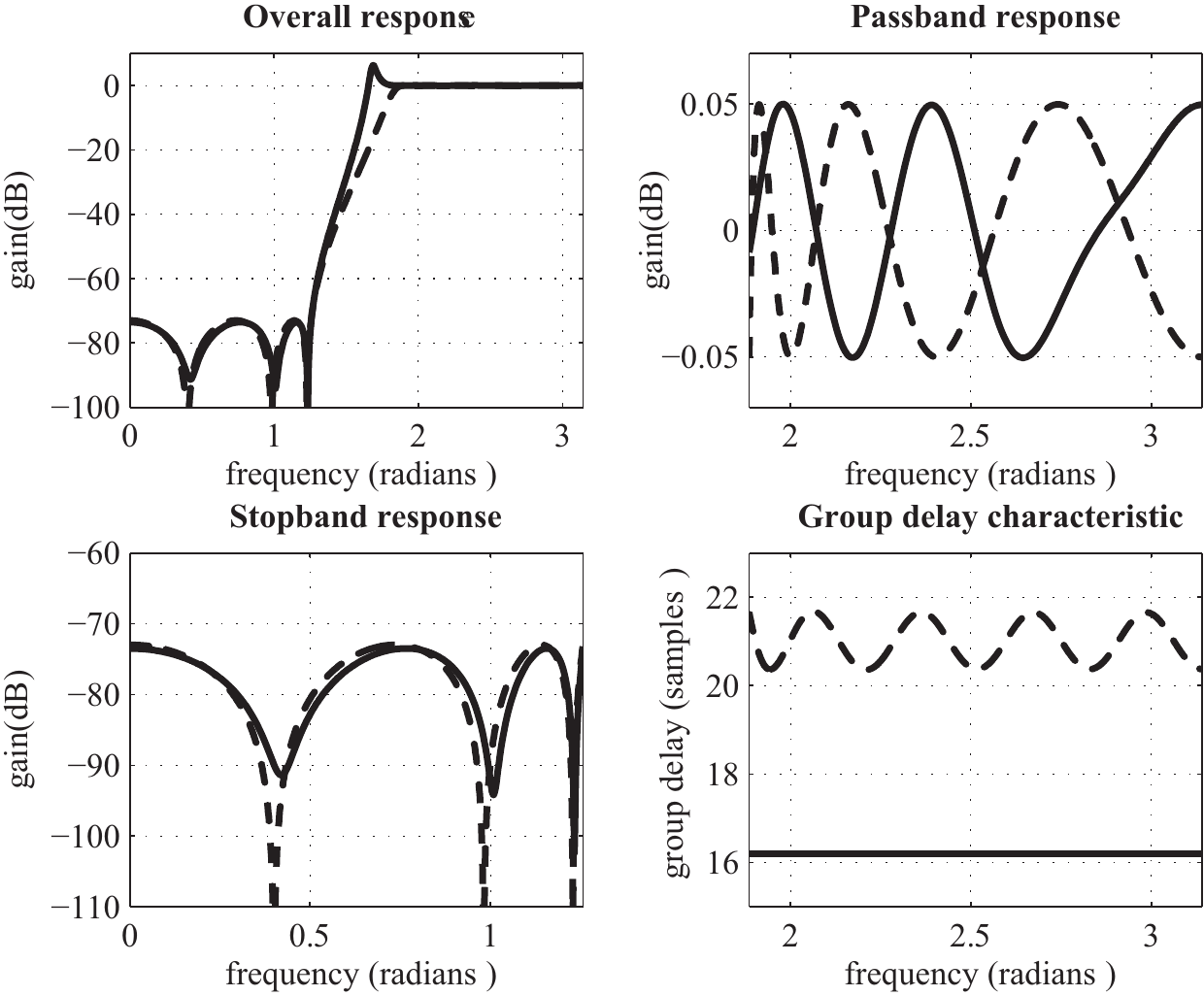}
\caption{Overall, passband, and stopband amplitude responses and group-delay characteristic for Design A-1 of the proposed method (solid curves) and the method in~\cite{andreasPhaseEq} (dashed curves) for Example 2.}
\label{eg2}
\end{center}
\end{figure}

\begin{figure}
\begin{center}
\includegraphics[width=0.48\textwidth]{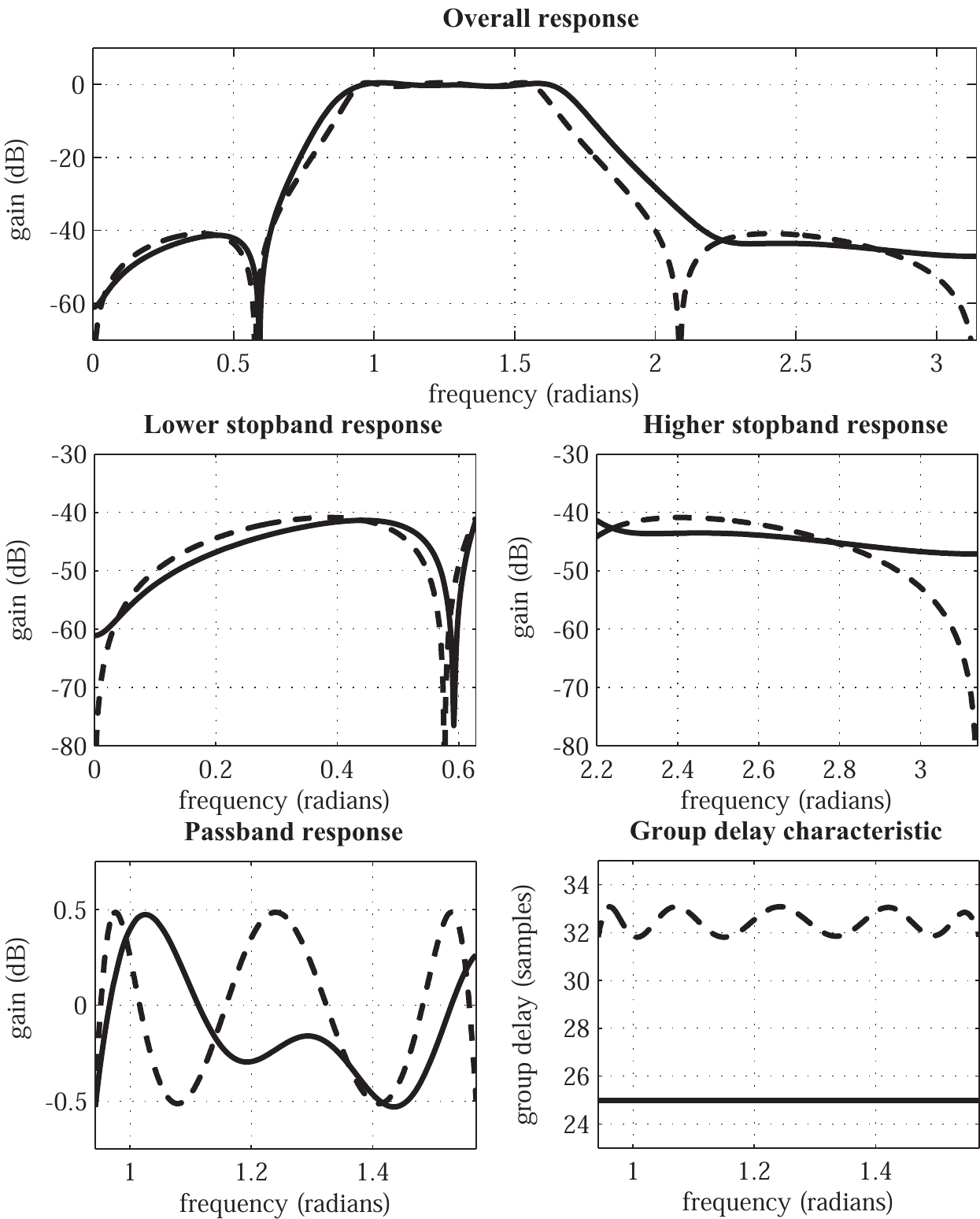}
\caption{Overall, stopband, and passband amplitude responses and group-delay characteristic for Design A-1 of the proposed method (solid curves) and the method in~\cite{andreas} (dashed curves) for Example 3.}
\label{eg3}
\end{center}
\end{figure}

The required design specifications for the filters of Examples 1, 2, and 3 are given in Tables~\ref{tab_eg1_specs}, \ref{tab_eg2_specs}, and \ref{tab_eg3_specs}, the results obtained are summarized in Tables~\ref{tab_eg1_results}, ~\ref{tab_eg2_results}, and ~\ref{tab_eg3_results} and the  amplitude responses and delay characteristics are plotted in Figs.~\ref{eg1}, \ref{eg2}, and \ref{eg3}, respectively. The poles and zeros of our designs for Examples 1 to 3 and also 4 to 9 are given in \cite{PolesZeros}. In designs where a transition-band anomaly occurred both designs A-1 and A-2 were carried out as seen in the tables.

The competing filters for Examples 1 and 2 correspond to the first and second examples in~\cite{andreasPhaseEq}, while that of Example 3 is obtained from p.~761 of~\cite{andreas}.
\begin{table}
\begin{center}
\caption{Lowpass Digital Filter Specifications for Example 1}
\label{tab_eg1_specs}
{\footnotesize{
\begin{tabular}{||l|c||} \hline \hline
Parameters & Values \\ \hline
Maximum passband ripple, dB  & 0.2 \\
Minimum stopband attenuation, dB  & 50  \\
Passband edge, rad/s  & $0.36\pi$  \\
Stopband edge, rad/s  & $0.44\pi$  \\
Maximum pole radius  & 0.98 \\
\hline \hline
\end{tabular}
}}
\end{center}
\end{table}

\begin{table}
\begin{center}
\caption{Design Results for Example 1 (Lowpass Filter)}
\label{tab_eg1_results}
{\footnotesize{
\begin{tabular}{||l|c|c|c||} \hline \hline
Parameters  & Design A-1 & Design A-2 & Method \\
 &  & $\Gamma_{tb} = 1 $  & in~\cite{andreasPhaseEq} \\
\hline
Elliptic filter order  & NA & NA & 6 \\
Equalizer filter order & NA & NA & 10 \\
Total filter order & 16 & 16  & 16 \\
Max PB ripple, dB & 0.2 & 0.2 & 0.2 \\
Min SB atten., dB  & 50.14 & 50.14 & 50.14\\
Max TB gain, dB& 2.5 & 0  & -0.2  \\
$\tau_{avg}$  & 24.38 & 24.45 & 29.75 \\
$Q_{\tau}$  & 0.00796 & 0.0132  & 6.82 \\
\hline \hline
\end{tabular}
}}
\\ PB: passband; TB:transition band; SB: stopband
\end{center}
\end{table}

\begin{table}
\begin{center}
\caption{Highpass Digital Filter Specifications for Example 2}
\label{tab_eg2_specs}
{\footnotesize{
\begin{tabular}{||l|c||} \hline \hline
Parameters & Values \\ \hline
Maximum passband ripple, dB  & 0.1 \\
Minimum stopband attenuation, dB  & 73  \\
Passband edge, rad/s  & $0.6\pi$  \\
Stopband edge, rad/s  & $0.4\pi$  \\
Maximum pole radius  & 0.98 \\
\hline \hline
\end{tabular}
}}
\end{center}
\end{table}

\begin{table}
\begin{center}
\caption{Design Results for Example 2 (Highpass Filter)}
\label{tab_eg2_results}
{\footnotesize{
\begin{tabular}{||l|c|c|c||} \hline \hline
Parameters  & Design A-1 & Design A-2 & Method \\
  &  & $\Gamma_{tb} = 1 $ & in~\cite{andreasPhaseEq} \\
\hline
Elliptic filter order  & NA & NA & 6 \\
Equalizer filter order  & NA & NA & 8 \\
Total filter order  & 14 & 14 & 14 \\
Max PB ripple, dB  & 0.1 & 0.1 & 0.1 \\
Min SB atten., dB  & 73.5 & 73.5 & 73.1 \\
Max TB gain, dB & 6.32 & 0.002  & -0.103 \\
$\tau_{avg}$  & 16.19 & 18.02 & 21 \\
$Q_{\tau}$ & 0.00104 & 0.00905 & 3.13\\
\hline \hline
\end{tabular}
}}
\\ PB: passband; TB:transition band; SB: stopband
\end{center}
\end{table}

\begin{table}
\begin{center}
\caption{Bandpass Digital Filter Specifications for Example 3}
\label{tab_eg3_specs}
{\footnotesize{
\begin{tabular}{||l|c||} \hline \hline
Parameters & Values \\ \hline
Maximum passband ripple, dB  & 1.0 \\
Minimum stopband attenuation, dB  & 41  \\
Low stopband edge, rad/s  & $0.2\pi$  \\
Low passband edge, rad/s  & $0.3\pi$  \\
High passband edge, rad/s  & $0.5\pi$  \\
High stopband edge, rad/s  & $0.7\pi$  \\
Maximum pole radius  & 0.98 \\
\hline \hline
\end{tabular}
}}
\end{center}
\end{table}

\begin{table}
\begin{center}
\caption{Design Results for Example 3 (Bandpass Filter)}
\label{tab_eg3_results}
{\footnotesize{
\begin{tabular}{||l|c|c|c||} \hline \hline
Parameters & Design A-1 & Design A-2 & Method  \\
  &  & $\Gamma_{tb} = 1 $ & in~\cite{andreas} \\
\hline
Elliptic filter order  & NA & NA & 6 \\
Equalizer filter order  & NA & NA & 8 \\
Total filter order  & 14 & 14 & 14 \\
Max PB ripple, dB  & 1 & 1 & 1 \\
Min SB atten., dB  & 41.4 & 41.37 & 41.37 \\
Max TB gain, dB  & 0.33 & 0 & -1 \\
$\tau_{avg}$  & 24.93 & 25.54 & 32.44 \\
$Q_{\tau}$  & 0.000461 & 0.00126 & 1.96 \\
\hline \hline
\end{tabular}
}}
\\ PB: passband; TB:transition band; SB: stopband
\end{center}
\end{table}
As can be seen in Tables~\ref{tab_eg1_results}, ~\ref{tab_eg2_results}, and ~\ref{tab_eg3_results}  and Figs.~\ref{eg1}, \ref{eg2}, and \ref{eg3}, the  IIR filters designed using the proposed method have much smaller maximum group-delay deviation for practically the same passband ripple and minimum stopband attenuation as the designs obtained with the classical equalizer method.

\subsection{Examples 4, 5, 6, 7, and 8}
The competing filters for Examples 4, 5, and 6 correspond to the fourth example in~\cite{guindon}, first example in~\cite{hinamoto}, and first example in~\cite{lang}, respectively. For Example 7 there are two competing filters, namely, the second example in~\cite{jiang} and the fourth example in~\cite{jiang2}, while for Example 8 the two competing filters were taken from the third example in~\cite{lai1}.

For Examples 5, 7, and 8 the group delay was prescribed by the competing methods and, therefore, both designs A and B were carried out.

In Example 6, we considered the design of a lowpass filter following a strategy reported in~\cite{lang} whereby the filter is designed as a cascade combination of an IIR filter and an FIR filter. For the comparison, we considered two filter structures for the proposed method: the first is similar to the structure used in~\cite{lang} and comprises a 4th-order IIR filter in cascade with an 11th-order FIR filter; the second is a 10th-order IIR filter which is equivalent to the first structure in terms of the number of multipliers.

The prescribed specifications for the filters in Examples 4 to 8 are given in Tables VII, IX, XII, XV, and XVIII, respectively. The results obtained are summarized in Tables VIII, X, XI, XIII, XIV, XVI, XVII, XIX, and XX and the frequency responses obtained are plotted in Figs. 4, 5, and 6. As can be seen in these tables and figures,
the  IIR filters designed using the proposed method have much smaller maximum group-delay deviation for practically the same passband ripple and minimum stopband attenuation as the designs obtained with the competing methods in~\cite{guindon}, \cite{hinamoto}, \cite{lang}, \cite{jiang}, \cite{jiang2}, and \cite{lai1} except that our method also yields increased minimum stopband attenuation relative to that achieved with the method in~\cite{lang}.

\vfill

\begin{table}
\begin{center}
\caption{Lowpass Digital Filter Specifications for Example 4}
\label{tab_eg4x_specs}
{\footnotesize{
\begin{tabular}{||l|c||} \hline \hline
Parameters & Values \\ \hline
Maximum passband ripple, dB  & 0.025 \\
Minimum stopband attenuation, dB  & 50  \\
Passband edge, rad/s  & $0.4\pi$  \\
Stopband edge, rad/s  & $0.6\pi$  \\
Maximum pole radius  & 0.98 \\
\hline \hline
\end{tabular}
}}
\end{center}
\end{table}

\begin{table}
\begin{center}
\caption{Design Results for Example 4 (Lowpass Filter)}
\label{tab_eg4x_results}
{\footnotesize{
\begin{tabular}{||l|c|c|c||} \hline \hline
Parameters & Design A-1 & Design A-2 & Method \\
 &  & $\Gamma_{tb} = 1 $  & in~\cite{guindon} \\
\hline
Total filter order & 10 & 10 & 10 \\
Max PB ripple, dB & 0.022 & 0.022 & 0.023 \\
Min SB atten., dB & 51 & 51  & 50.9 \\
Max TB gain, dB & 10.18 & 0.002 & 0.17 \\
$\tau_{avg}$ & 8.35  & 8.59 & 9.28 \\
$Q_{\tau}$ & 0.000472 & 0.20 & 1.07 \\
\hline \hline
\end{tabular}
}}
\\ PB: passband; TB:transition band; SB: stopband
\end{center}
\end{table}
\begin{figure}
\begin{center}
\includegraphics[width=0.48\textwidth]{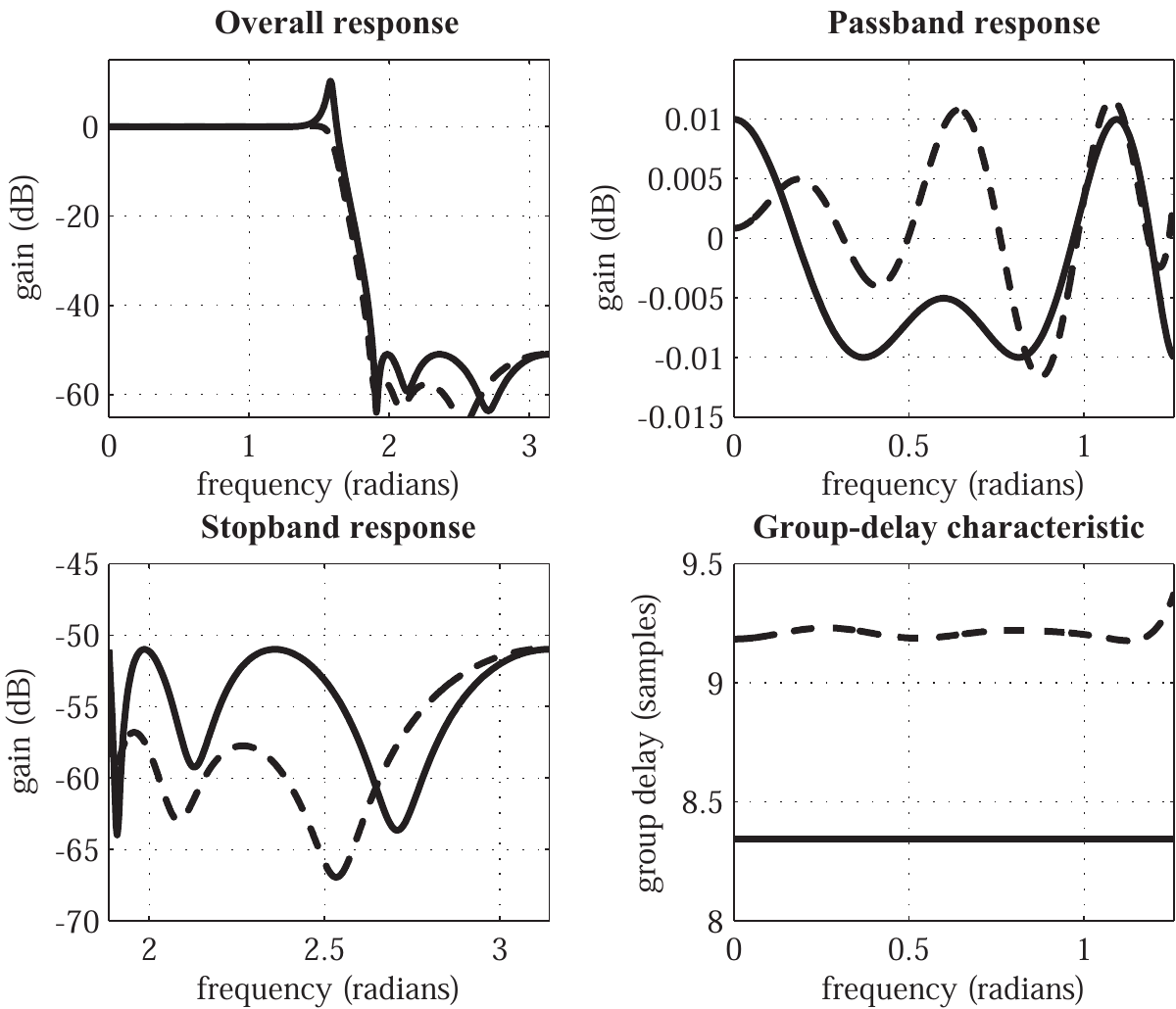}
\caption{Overall, passband, and stopband amplitude responses and group-delay characteristic for Design A-1 of the proposed method (solid curves) and the method in~\cite{guindon} (dashed curves) for Example 4.}
\label{eg4x}
\end{center}
\end{figure}

\begin{table}
\begin{center}
\caption{Lowpass Digital Filter Specifications for Example 5}
\label{tab_eg4_specs}
{\footnotesize{
\begin{tabular}{||l|c||} \hline \hline
Parameters & Values \\ \hline
Maximum passband ripple, dB  & 0.266 \\
Minimum stopband attenuation, dB  & 37  \\
Passband edge, rad/s  & $0.5\pi$  \\
Stopband edge, rad/s  & $0.6\pi$  \\
Maximum pole radius  & 0.98 \\
Group delay (samples) & 15.9 \\
\hline \hline
\end{tabular}
}}
\end{center}
\end{table}

\begin{table}
\begin{center}
\caption{Design Results for Example 5 for Variable Group Delay (Lowpass filter)}
\label{tab_eg4a_results}
{\footnotesize{
\begin{tabular}{||l|c|c|c||} \hline \hline
Parameters  & Design A-1 & Design A-2 & Method \\
  &  & $\Gamma_{tb} = 1 $ & in~\cite{hinamoto} \\
\hline
Total filter order  & 12 & 12 & 12 \\
Max PB ripple, dB  & 0.265 & 0.265 & 0.266 \\
Min SB atten., dB  & 36.145  & 36.146 & 36.146 \\
Max TB gain, dB  & 4.37 & 0 & -0.136 \\
$\tau_{avg}$  & 10.92 & 11.40 & 16.26  \\
$Q_{\tau}$  & 0.00449 & 0.0188 & 4.54 \\
\hline \hline
\end{tabular}
}}
\\ PB: passband; TB:transition band; SB: stopband
\end{center}
\end{table}
\begin{table}
\begin{center}
\caption{Design Results for Example 5 for Fixed Group Delay (Lowpass Filter)}
\label{tab_eg4b_results}
{\footnotesize{
\begin{tabular}{||l|c|c||} \hline \hline
Parameters  & Design B-1 & Method \\
  &   & in~\cite{hinamoto} \\
\hline
Total filter order & 12 & 12  \\
Max PB ripple, dB & 0.263 & 0.266 \\
Min SB atten., dB & 36.146 & 36.146   \\
Max TB gain, dB  & -0.023 & -0.136 \\
$\tau_{avg}$   &  15.98 & 16.26 \\
$Q_{\tau}$  & 2.69 & 4.54 \\
\hline \hline
\end{tabular}
}}
\\ PB: passband; TB:transition band; SB: stopband
\end{center}
\end{table}
\begin{figure}
\begin{center}
\includegraphics[width=0.48\textwidth]{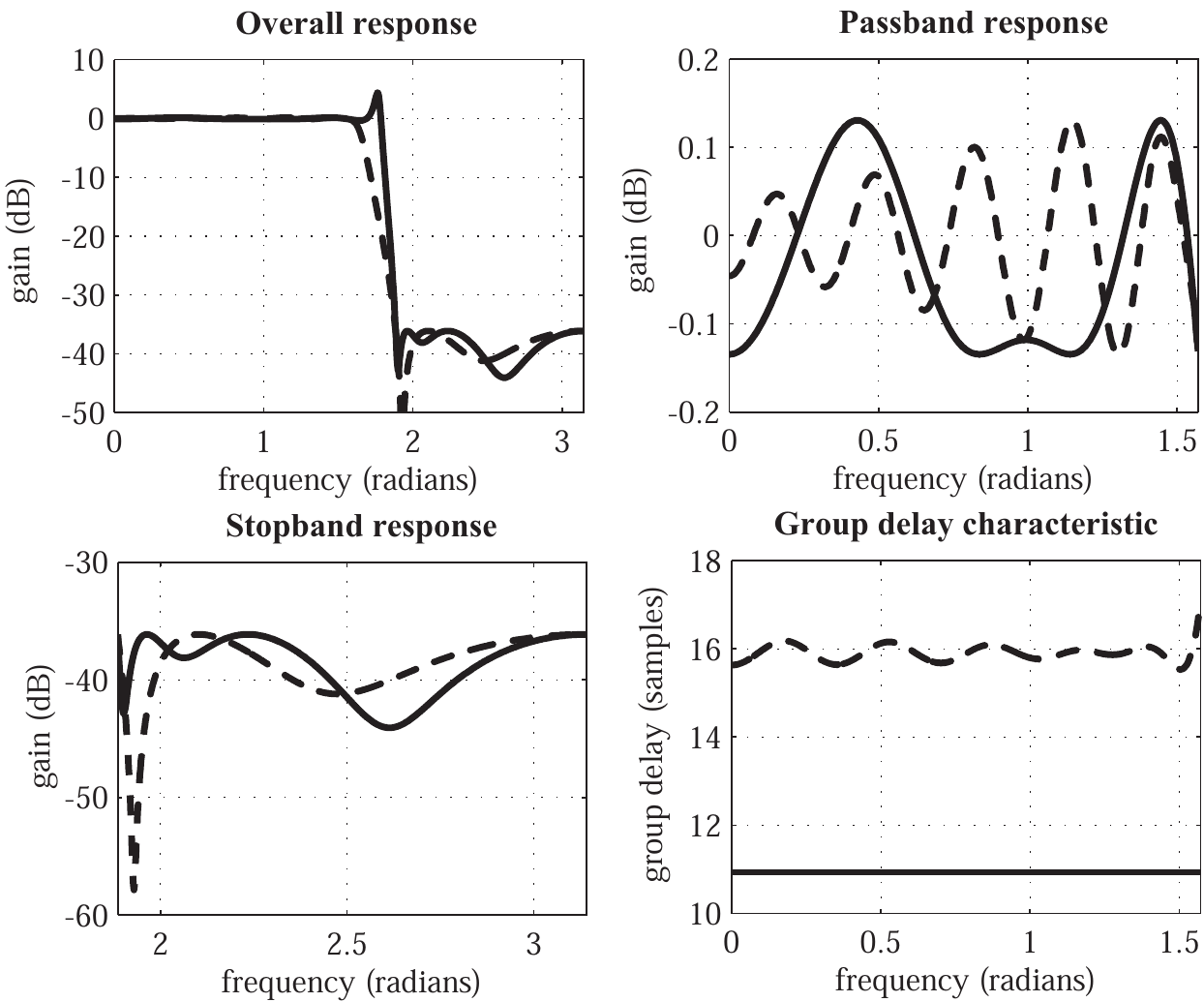}
\caption{Overall, passband, and stopband amplitude responses and group-delay characteristic for Design A-1 of the proposed method (solid curves) and the method in~\cite{hinamoto} (dashed curves) for Example 5.}
\label{eg4}
\end{center}
\end{figure}
\begin{table}
\begin{center}
\caption{Lowpass Digital Filter Specifications for Example 6}
\label{tab_eg5_specs}
{\footnotesize{
\begin{tabular}{||l|c||} \hline \hline
Parameters & Values \\ \hline
Maximum passband ripple, dB  & 0.25 \\
Minimum stopband attenuation, dB  & 44  \\
Passband edge, rad/s  & $0.4\pi$  \\
Stopband edge, rad/s  & $0.56\pi$  \\
Maximum pole radius  & 0.98 \\
\hline \hline
\end{tabular}
}}
\end{center}
\end{table}
\begin{table}
\begin{center}
\caption{Design Results for Example 6 Where the Proposed Filter Is an All-IIR Filter Structure (Lowpass Filter)}
\label{tab_eg5a_results}
{\footnotesize{
\begin{tabular}{||l|c|c|c||} \hline \hline
Parameters  & Design A-1 & Design A-2 & Method \\
 &  & $\Gamma_{tb} = 1 $  & in~\cite{lang} \\
\hline
IIR filter order  & 10 & 10 & 4 \\
FIR filter order  & 0 & 0 & 11 \\
Max PB ripple, dB  & 0.207  & 0.207 & 0.207 \\
Min SB atten., dB  & 50 & 50 & 44.1 \\
Max TB gain, dB  & 7.89 & 0 & -0.133 \\
$\tau_{avg}$  & 9.15 & 9.79 & 12.2 \\
$Q_{\tau}$  & 0.00130 & 0.159 & 2.25 \\
\hline \hline
\end{tabular}
}}
\\ PB: passband; TB:transition band; SB: stopband
\end{center}
\end{table}
\begin{table}[ht]
\begin{center}
\caption{Design Results for Example 6 Where the Proposed Filter Is an IIR-FIR Filter Combination (Lowpass Filter)}
\label{tab_eg5b_results}
{\footnotesize{
\begin{tabular}{||l|c|c|c||} \hline \hline
Parameters  & Design A-1 & Design A-2 & Method \\
  &  & $\Gamma_{tb} = 1 $ & in~\cite{lang} \\
\hline
IIR filter order & 4 & 4 & 4 \\
FIR filter order & 11 & 11 & 11 \\
Max PB ripple, dB  & 0.204  & 0.205 & 0.207 \\
Min SB atten., dB  & 50 & 50 & 44.1 \\
Max TB gain, dB & 8.77 & 0 & -0.133 \\
$\tau_{avg}$  & 8.72 & 9.35 & 12.2 \\
$Q_{\tau}$  & 0.0149 & 0.382 & 2.25 \\
\hline \hline
\end{tabular}
}}
\\ PB: passband; TB:transition band; SB: stopband
\end{center}
\end{table}
\begin{figure}
\begin{center}
\includegraphics[width=0.48\textwidth]{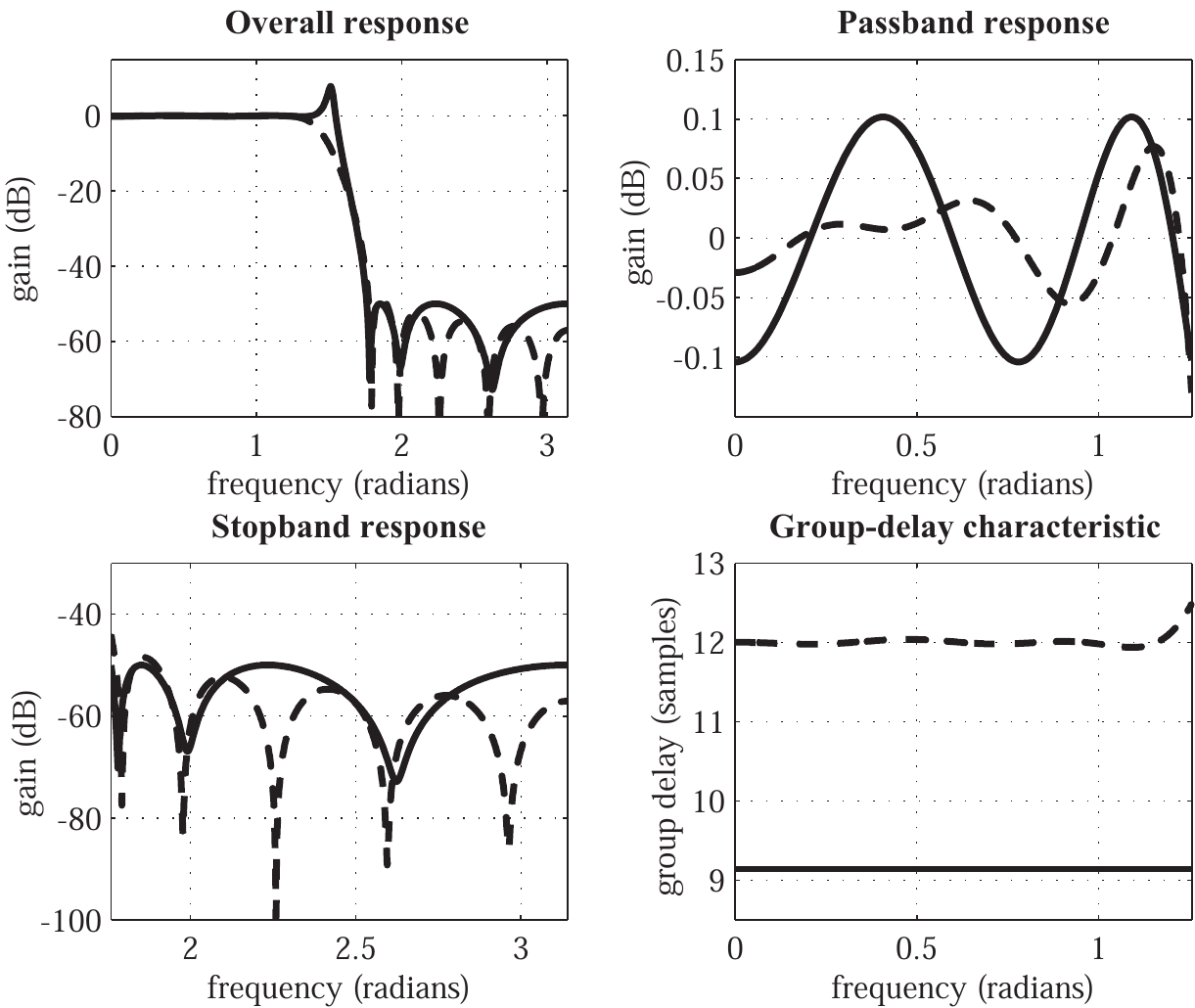}
\caption{Overall, passband, and stopband amplitude responses and group-delay characteristic for Design A-1 of the all-IIR filter structure of the proposed method (solid curves) and the method in~\cite{lang} (dashed curves) for Example 6.}
\label{eg5}
\end{center}
\end{figure}

\begin{table}
\begin{center}
\caption{Highpass Digital Filter Specifications for Example 7}
\label{tab_eg6_specs}
{\footnotesize{
\begin{tabular}{||l|c||} \hline \hline
Parameters & Values \\ \hline
Maximum passband ripple, dB  & 0.720 \\
Minimum stopband attenuation, dB  & 27  \\
Passband edge, rad/s  & $0.525\pi$  \\
Stopband edge, rad/s  & $0.475\pi$  \\
Maximum pole radius  & 0.98 \\
Group delay (samples) & 12 \\
\hline \hline
\end{tabular}
}}
\end{center}
\end{table}
\begin{table}
\begin{center}
\caption{Design Results for Example 7 for Variable Group Delay (Highpass Filter)}
\label{tab_eg6a_results}
{\footnotesize{
\begin{tabular}{||l|c|c|c||} \hline \hline
Parameters  & Design A-1 & Design A-2 & Method\\
  &   & $\Gamma_{tb} = 1$ & in~\cite{jiang}   \\
\hline
Filter order & 14 & 14 & 14 \\
Max PB ripple, dB  & 0.491 & 0.492 & 0.720  \\
Min SB atten., dB  & 31 & 30.98 & 27.36  \\
Max TB gain, dB  & 7.5 & 0.021 & 0.155  \\
$\tau_{avg}$  & 11.95 & 12.7 & 13.74  \\
$Q_{\tau}$  & 0.00515 & 1.71 & 16.4  \\
\hline
Parameters  & Design A-1 & Design A-2 &  Method  \\
  &   & $\Gamma_{tb} = 1$ &  in~\cite{jiang2}  \\
\hline
Filter order & 14 & 14 &  14\\
Max PB ripple, dB  & 0.491 & 0.492 & 0.492 \\
Min SB atten., dB  & 31 & 30.98 &  30.7 \\
Max TB gain, dB  & 7.5 & 0.021 &  3.86  \\
$\tau_{avg}$  & 11.95 & 12.7 &  12.79 \\
$Q_{\tau}$  & 0.00515 & 1.71 & 9 \\
\hline \hline
\end{tabular}
}}
\\ PB: passband; TB:transition band; SB: stopband
\end{center}
\end{table}
\begin{table}
\begin{center}
\caption{Design Results for Example 7 for Fixed Group Delay (Highpass Filter)}
\label{tab_eg6b_results}
{\footnotesize{
\begin{tabular}{||l|c|c|c||} \hline \hline
Parameters  & Design B-1 & Design B-2 & Method\\
  &  & $\Gamma_{tb} = 1 $ &in~\cite{jiang}  \\
\hline
Filter order & 14 & 14 & 14 \\
Max PB ripple, dB  & 0.491 & 0.491 &  0.720  \\
Min SB atten., dB  & 31 & 31 &  27.36  \\
Max TB gain, dB  & 7.08 & 0.022 &  0.155  \\
$\tau_{avg}$  & 12 & 12 & 13.74  \\
$Q_{\tau}$  & 0.018 & 4.8 & 16.4  \\
\hline \hline
Parameters  & Design B-1 & Design B-3 &  Method \\
  &  & & in~\cite{jiang2} \\
\hline
Filter order & 14  & 14 & 14\\
Max PB ripple, dB  & 0.491  & 0.491 & 0.492 \\
Min SB atten., dB  & 31 & 31 &  30.7 \\
Max TB gain, dB  & 7.08  & 3.79 &  3.86 \\
$\tau_{avg}$  & 12 & 12 &  12.79 \\
$Q_{\tau}$  & 0.018 & 0.028 &  9 \\
\hline \hline
\end{tabular}
}}
\\ PB: passband; TB:transition band; SB: stopband
\end{center}
\end{table}
\begin{table}
\begin{center}
\caption{Lowpass Digital Filter Specifications for Example 8}
\label{tab_eg6X_specs}
{\footnotesize{
\begin{tabular}{||l|c||} \hline \hline
Parameters & Values \\ \hline
Maximum passband ripple, dB  & 0.1 \\
Minimum stopband attenuation, dB  & 44  \\
Passband edge, rad/s  & $0.5\pi$  \\
Stopband edge, rad/s  & $0.55\pi$  \\
Maximum pole radius  & 0.98 \\
Group delay (samples) & 15 \\
\hline \hline
\end{tabular}
}}
\end{center}
\end{table}
\begin{table}
\begin{center}
\caption{Design Results for Example 8 for Variable Group Delay (Lowpass Filter)}
\label{tab_eg6Xa_results}
{\footnotesize{
\begin{tabular}{||l|c|c|c||} \hline \hline
Parameters  & Design A-1 & Design A-2 & Method in~\cite{lai1}  \\
  &  & $\Gamma_{tb} = 1$ & 1st Design   \\
\hline
Filter order & 18 & 18 & 18  \\
Max PB ripple, dB  & 0.0904 & 0.09 & 0.0995  \\
Min SB atten., dB  & 45.5 & 45.5 & 44.77  \\
Max TB gain, dB  & 7 & 0 & 11.25  \\
$\tau_{avg}$  & 19.6 & 20.69 & 14.45 \\
$Q_{\tau}$  & 0.204 & 3.35 & 5.65  \\
\hline
Parameters  & Design A-1 & Design A-2 &  Method in~\cite{lai1} \\
  &  & $\Gamma_{tb} = 1$ & 2nd Design  \\
\hline
Filter order & 18 & 18 & 18 \\
Max PB ripple, dB  & 0.0904 & 0.09 &  0.0907 \\
Min SB atten., dB  & 45.5 & 45.5 & 45.36 \\
Max TB gain, dB  & 7 & 0 &  21.90  \\
$\tau_{avg}$  & 19.6 & 20.69 & 14.46 \\
$Q_{\tau}$  & 0.204 & 3.35 & 5.09 \\
\hline \hline
\end{tabular}
}}
\\ PB: passband; TB:transition band; SB: stopband
\end{center}
\end{table}
\begin{table}
\begin{center}
\caption{Design Results for Example 8 for Fixed Group Delay (Lowpass Filter)}
\label{tab_eg6Xb_results}
{\footnotesize{
\begin{tabular}{||l|c|c|c||} \hline \hline
Parameters  & Design B-1 & Method in~\cite{lai1} & Method in~\cite{lai1}  \\
  &  & 1st Design  & 2nd Design \\
\hline
Filter order & 18 & 18 & 18 \\
Max PB ripple, dB  & 0.09 & 0.0995 & 0.0907 \\
Min SB atten., dB  & 45.5 & 44.77 & 45.36  \\
Max TB gain, dB  & 7.43 & 11.25 & 21.90  \\
$\tau_{avg}$  & 15 & 14.45 &  14.46 \\
$Q_{\tau}$  & 3.18 & 5.65 & 5.09 \\
\hline \hline
\end{tabular}
}}
\\ PB: passband; TB:transition band; SB: stopband
\end{center}
\end{table}

\subsection{Example 9}
We also carried out comparisons with lowpass filter 1-iii in the first example of~\cite{sullivan}. Our designs A-1, A-2, and B-1 have slightly better amplitude-response specifications and, in addition, offer reduced  values of $Q_\tau$ relative to that achieved in~\cite{sullivan}, namely, 0.00043, 0.00089, 0.086 for designs A-1, A-2, and B-1, respectively, versus 0.13 for the design in~\cite{sullivan}. We then compared the same designs with filter MMPE(2) in the first example of~\cite{lai2} and again found that our designs offer slightly better amplitude-response specifications and much lower values of $Q_\tau$ relative to that for the design in~\cite{lai2} which was 0.11. The poles and zeros and additional results for these filters are given in~\cite{PolesZeros}. It should be noted that no filter coefficients are provided in~\cite{lai2} but from the design results reported in~\cite{lai2} we were able to estimate all the performance parameters except for the maximum transition-band gain.

\subsection{Example 10}
The prescribed specifications for the high-selectivity bandpass filter are given in Table XXI.  For this filter, we obtained type A-1 and A-2 IIR designs and also a corresponding optimal FIR design that satisfies the same specifications. The optimization was carried out for various IIR-filter orders by varying the number of additional biquadratic sections. The FIR filter was designed using the {\it weighted-Chebyshev} method described in Chapter 15 of~\cite{andreas}.
The results obtained are presented in Tables~\ref{tab_eg7a_results} and \ref{tab_eg7b_results}. In these tables, we also include the numbers of multiplications, additions, and unit delays per sampling period required by each design. These numerical values correspond to a metric of the computational effort required in a software implementation or the complexity of the hardware in a hardware implementation since the number of multiplications, additions, and unit delays translate directly into multipliers, adders, and unit-delay elements, respectively. We have assumed a cascade realization of SOSs both for the IIR and FIR filters. For the IIR filters we have assumed a direct canonic realization which would require a total of $4J+1$ multiplications, $4J$ additions, and $2J$ unit delays in general where $J$ is the number of sections~\cite{andreas}.   In the case of the FIR filter, $J+1$ multiplications, $2J$ additions, and $2J$ unit delays would be required in view of the symmetry property of the transfer function coefficients.
\begin{table}
\begin{center}
\caption{Bandpass Digital Filter Specifications for Example 10}
\label{tab_eg7_specs}
{\footnotesize{
\begin{tabular}{||l|c||} \hline \hline
Parameters & Values \\ \hline
Maximum passband ripple, dB  & 1.0 \\
Minimum stopband attenuation, dB  & 68  \\
Low stopband edge, rad/s  & 1  \\
Low passband edge, rad/s  & 1.1  \\
High passband edge, rad/s  & 1.9 \\
High stopband edge, rad/s  & 2  \\
Maximum pole radius  & 0.991 \\
\hline \hline
\end{tabular}
}}
\end{center}
\end{table}
From Tables~\ref{tab_eg7a_results} and \ref{tab_eg7b_results}, we observe a clear trade-off between filter complexity and group delay versus maximum group-delay deviation. Evidently, an IIR design would offer significant reduction in the number of arithmetic operations and group delay of the filter but at the cost of a higher maximum group-delay deviation. For most applications, a perfectly linear-phase response is not required and a value of $Q_\tau$ in the range of 1 to 10, depending on the application, would be entirely acceptable. In such applications, an IIR design that is much more economical than an FIR design would be possible. Note that the complexity of the IIR design is reduced and its efficiency is increased as the allowable value of $Q_\tau$ is increased. This constitutes a  most valuable trade-off that would allow a filter designer to design the most economical filter that would satisfy the required specifications for the intended application.
\begin{table*}[htb*]
\begin{center}
\caption{Comparison Between Design A-1 of the IIR Filters And an Equivalent FIR Filter (Example 10)}
\label{tab_eg7a_results}
{\footnotesize{
\begin{tabular}{||l|c|c|c|c|c|c||} \hline \hline
Parameters & Elliptic & IIR & IIR & IIR & IIR & FIR \\
 & Filter & Filter 1 & Filter 2 & Filter 3 & Filter 4 & Filter \\
\hline
Total filter order & 14 & 20 & 22 & 24 & 26 & 142 \\
Max PB ripple, dB & 1 & 0.98 & 0.98 & 0.98 & 0.78 & 0.98 \\
Min SB atten., dB & 68.5 & 68 & 68 & 68 & 68.3 & 68 \\
Max TB gain, dB & -1 & 11.73 & 10.71 & 11.36 & 10.64 & -0.5 \\
$\tau_{avg}$ & 63.33 & 29.5 & 34.1 & 38.57 & 41.72 & 71 \\
$Q_{\tau}$ & 83.77 & 19.85 & 6.95 & 1.89 & 0.58 & 0 \\
No. of multiplications & 29 & 41 & 45 & 49 & 53 & 72 \\
No. of additions & 28 & 40 & 44 & 48 & 52 & 142 \\
No of delays & 14 & 20 & 22 & 24 & 26 & 142 \\
\hline \hline
\end{tabular}
}}
\\ PB: passband; TB:transition band; SB: stopband
\end{center}
\end{table*}
\begin{table*}[htb*]
\begin{center}
\caption{Comparison Between Design A-2 of the IIR Filters And an Equivalent FIR Filter (Example 10)}
\label{tab_eg7b_results}
{\footnotesize{
\begin{tabular}{||l|c|c|c|c|c|c||} \hline \hline
Parameters & Elliptic & IIR & IIR & IIR & IIR & FIR \\
 & Filter & Filter 1 & Filter 2 & Filter 3 & Filter 4 & Filter   \\
\hline
Total filter order & 14 & 22 & 24 & 26 & 28 & 142  \\
Max PB ripple, dB & 1 & 0.98 & 0.98 & 0.98 & 0.98 & 0.98 \\
Min SB atten., dB & 68.5 & 68 & 68 & 68 & 68 & 68 \\
Max TB gain, dB & -1 & 0.01 & 0.01 & 0.02 & 0.01 & -0.5 \\
$\tau_{avg}$ & 63.33 & 33.66 & 38.59 & 44.75 & 49.27 & 71 \\
$Q_{\tau}$ & 83.77 & 17.38 & 9.6 & 3.6 & 0.49 & 0 \\
No. of multiplications & 29 & 45 & 49 & 53 & 57 & 72 \\
No. of additions & 28 & 44 & 48 & 52 & 56 & 142 \\
No of delays & 14 & 22 & 24 & 26 & 28 & 142 \\
\hline \hline
\end{tabular}
}}
\\ PB: passband; TB:transition band; SB: stopband
\end{center}
\end{table*}

The above design examples have shown that the proposed design method yields filters that satisfy arbitrary prescribed amplitude-response specifications with the lowest maximum group-delay deviation compared to those achieved with the competing design methods considered. The solution obtained sometimes depends on the initialization filter and group delay used, which implies that global convergence cannot be guaranteed.  In effect, the proposed method gives quality suboptimal designs that may sometimes be globally optimal.

\section{Conclusion}
A method for the design of nearly linear-phase IIR digital filters that satisfies prescribed specifications has been described. In the proposed method, the group-delay deviation is minimized under the constraints that the passband ripple and minimum stopband attenuation meet the specifications and both a prescribed or an optimized group delay can be achieved. By designing the filter as a cascade of second-order sections, a nonrestrictive stability constraint characterized by a set of linear inequality constraints can be incorporated in the optimization algorithm. An additional feature of the method, which is very useful in certain applications, is the inherent capability of constraining the maximum gain in the transition band to be below a prescribed value. This facilitates the elimination of transition-band anomalies which sometimes occur in filters designed by optimization.

Experimental results have shown that the nearly linear-phase IIR filters designed using the proposed method have a much lower maximum group-delay deviation for the same passband ripple and minimum stopband attenuation when compared with several filters designed with state-of-the-art competing methods. It has also been demonstrated that nearly linear-phase IIR filters offer some substantial advantages when compared with their exact linear-phase FIR counterparts such as lower group delay and filter complexity without compromising the required amplitude-response specifications.

\section*{Acknowledgment}
The authors are grateful to the Natural Sciences and Engineering Research Council of Canada for supporting this work.

\ifCLASSOPTIONcaptionsoff
  \newpage
\fi



%

\end{document}